\documentclass[aps,prl,preprint,superscriptaddress,amsmath,amssymb,showpacs]{revtex4}
\usepackage{hyperref}
\hypersetup{colorlinks=black}
\draft % marks overfull lines with a black rule on the right
\usepackage{graphicx}
\begin{document}

%preprint{Applied Physics Letters}

\title{\textbf{Electron Thermionic Emission from Graphene and Thermionic Energy Converter}} %Title of paper



\author{Shi-Jun Liang}
\affiliation{SUTD-MIT International Design Center, Singapore University of Technology and Design, Singapore 138682.}
\author{L. K. Ang}
\email{ricky\_ang@sutd.edu.sg}
\affiliation{SUTD-MIT International Design Center, Singapore University of Technology and Design, Singapore 138682.}
\date{\today}

\begin{abstract}
In this paper, we propose a model to investigate the electron thermionic emission from a single-layer graphene (ignoring the effects of substrate) and to explore its application as the emitter of thermionic energy convertor (TIC).
An analytical formula has been derived, which is a function of temperature, work function and Fermi energy level.
The formula is significantly different from the traditional Richardson-Dushman (RD) law for which  it is independent of mass to account for the supply function of the electrons in the graphene behaving like massless Fermion quasiparticles.
By comparing with a recent experiment [Kaili Jiang et al., Nano Research 7, 553 (2014)] measuring electron thermionic emission from a suspended single layer graphene, our model predicts that the intrinsic work function of a single-layer graphene is about 4.514 eV with a Fermi energy level of 0.083 eV.
For a given work function, a new scaling of $T^{3}$ is predicted, which is different from the traditional RD scaling of $T^2$.
If the work function of the graphene is lowered to 2.5 to 3 eV, and the Fermi energy level is increased to 0.8 to 0.9 eV,  it is possible to design a graphene cathode based TIC operating at around 900 K or lower, as compared with the metal-based cathode TIC (operating at about 1500 K).
With a graphene based cathode (work function = 4.514 eV) at 900 K, and a metallic based anode (work function = 2.5 eV) like LaB$_6$ at 425 K, the efficiency of our proposed-TIC is about 45$\%$.
\end{abstract}

\maketitle %\maketitle must follow title, authors, abstract and \pacs
%\section*{Project}

\section{Introduction}
Thermionic emission describes electrons evaporation from a heated cathode when the electrons gain sufficient energy from the thermal energy to overcome the potential barrier (or work function) near the cathode surface (see Fig. 1).
The amount of current density $J$ from the thermionic emission is determined by the Richardson-Dunshman (RD) law \cite{Richardson2003}, given by
\begin{equation}
J=AT^2exp[-\frac{\Phi}{k_{B}T}].\label{1}
\end{equation}
Here, $T$ is the cathode temperature, $\Phi$ is the work function of metal (independent of $T$), $k_{B}$ is the Boltzmann constant, the prefactor $A=4\pi e m k_{B}^{2}/h^{3}=1.2 \times10^6$ A m$^{-2}$ K$^{-2}$ is the Richardson constant, $e$ is the electron charge, and $m$ is the electron mass.
In general, RD law only works well for metallic-like materials, and a corresponding modification by using quantum models is required for wide band-gap materials and low electron affinity materials \cite{walker2013}.

Since the mono-layer graphene \cite{Novoselov2004} was exfoliated experimentally in 2004, many unique properties have been reported, such as linear band structure \cite{Novoselov2005}, ultra-high mobility (up to 40000 cm$\cdot$V$^{-1}$$\cdot$s$^{-1}$), and excellent conductivity \cite{Balandin2008}.
Fundamentally, the linear band structure of graphene, the most intriguing property, makes it different from other 3-dimensional or bulk materials.
For electron emission, it has been recently shown that the traditional emission processes, such as field emission and photo-assisted over-barrier electron emission, may require further revisions \cite{Sun2011,shijuna,SJLiang2014} to account for the unique properties of graphene. As the crystalline allotrope of carbon, the thermionic and field emission from carbon nanotube (CNT) had also been studied both experimentally \cite{WaltAdeHeer1995,JeanMarcBonard2002,Khondaker2012,Otto2003} and theoretically \cite{liangshidong2006,liangshidong2008,liangshidong2010},
which indicated that traditional emission models may not be valid for CNT.
Recent experiments\cite{xianlongwei2014,xianlongwei2011} have confirmed that RD law is not valid for thermionic emission from CNT.

In this paper, we are interested to know if the RD law (Eq. \ref{1}) is valid for thermionic emission from a single-layer suspended graphene by assuming that the effect of the substrate is not important.
Electrons in graphene behave as massless quasiparticles ($m$ = 0), so the mass-dependent expression of the prefactor $A$ in the RD law [Eq. (1)] is questionable as the supply function of the electron behaving like massless particles in graphene is not included.
On the other hand, the electrons in graphene must exhibit a non-zero mass when they are collectively excited, which is only a few percents (0.01 to 0.03) of the intrinsic electron mass \cite{Yoon2014}.

Harvesting thermal energy is important to maintain a sustainable energy needs since nearly 60\% of the energy input to our society is wasted as heat.
One of the most common methods in converting heat into electricity is based on thermoelectric (TE) materials, which has progressed significantly since 1990s by using low-dimensional materials  \cite{Zebarjadi2012}.
One of the limitation of TE power generators is low efficiency ($<$ 40\%) in the temperature ranges of 600 to 1000 K (see Fig. 6 in Ref. \cite{Zebarjadi2012}).
Another method for the conversion of heat to electricity is known as the thermionic energy convertor (TIC), for which electrons are evaporated from a heated cathode into a vacuum and then condensed at a cooler anode \cite{Hatsopoulos1974}.
Compared with TE based convertor (TEC), TIC normally operates at high temperature (above 1500 K) with higher efficiency ($>$ 50\%).
However, it is difficult to operate TIC at low temperature, as common metallic cathodes having  high work function cannot produce sufficient electron emission at low temperature.

The other limiting factor of TIC is the space charge effects of the electrons traveling across the gap, which have been addressed either by reducing the gap spacing or by using Cs positive ions.
Recently, some satisfactory progresses have been achieved to avoid this space charge problem by modifying the electric potential inside the gap \cite{Meir2013} or by introducing heterostructures  \cite{Jared2013}.

It is desirable if TIC can operate with relatively high efficiency (compared to TEC) at immediate temperature ranging from 700 K to 1000 K, which may \emph{significantly} complement the ongoing research of TEC and also being alternative energy harvesting devices in this intermediate temperature range. Some recent approaches to improving the efficiency of TIC are achieved by using photo-enhanced thermionic emission from a wide-band gap emitter \cite{Jared2010}, engineering surface effects \cite{Segev2013} and seeking alternative emitters \cite{Starodub2012}.

Other than studying the validity of using RD law to describe the electron thermionic emission from graphene, we are also interested in exploring if a graphene-based cathode TIC can have better efficiency than a metal-based cathode TIC at similar operating condition like same work function. From our model, we first derive an analytic solution [Eq.(\ref{55})] for the thermionic emission from graphene, which indicates that the traditional RD law has failed to describe thermionic emission from a single suspended layer of graphene [see Fig. 2(a)].
The calculated results are in a very good agreement with a very recent experimental data \cite{jiangkaili2014} [see Fig. 4(a)].
If the property of graphene can be tuned to some favorable parameters, we also predict that the proposed graphene cathode based TIC can have higher current density ($>$ 10 A/m$^2$) at 700 to 1000 K  [see Fig. 3].
Finally, we show that a graphene based cathode (work function = 4.514 eV) at 900 K, and a metallic based anode (work function = 2.5 eV) like LaB$_6$ at 425 K, the efficieny of our proposed TIC is about 45 $\%$ [see Fig. 5].

\section{Theoretical model}
%\subsection{The electron theory of graphene}

In Fig. 1a, we show the schematic diagram of our model.
By assuming that the graphene based cathode has been heated up to a uniform temperature $T_c$ (without considering the heating process and uniformity issue for simplicity), the electrons will be emitted from the cathode to reach the anode biased at temperature $T_a$, by overcoming the potential barrier as shown in Fig. 1b. The work function of the cathode and of anode is, respectively, defined as $\Phi_c$ and $\Phi_a$.
The amount of the emitted current density from the cathode and anode is, respectively $J_c$ and $J_a$, which will be calculated in this paper.

The electron theory of graphene proposed by Wallace \cite{Wallace1974} can provide the basis for practically all properties of graphene.
This theory implies that electron in the graphene mimics the Dirac fermion, and that the equation of electron motion obeys the 2D Dirac-like equation.
By using this model, the electron state is described by a two-component wave function for which the parallel electron energy is $E_p=\hbar v_{f}\mid \textbf{k}\mid$ in the low energy regime.
The number of electron states per unit cell with energy between $E_p$ and $E_p+dE_p$ is given by \cite{Castroneto2009}
\begin{equation}
  \rho(E_p)dE_p=\frac{2}{(2\pi)^2}\int\int d^2k=\frac{2E_p}{\pi(\hbar v_{f})^2}dE_p,
\end{equation}
where $\hbar$ is the reduced Planck constant and $v_{f}$ ($10^6$ m/s) is the velocity of massless Dirac fermions in the graphene.
The probability of an electron state with total energy $E$ being occupied is given by the Fermi-Dirac (FD) distribution function,

\begin{equation}
f_{F-D}(E)=\frac{1}{1+exp[(E-E_{f})/k_{B}T]},
\end{equation}
where  $E_f$ is the Fermi energy level (= 0.083 eV for intrinsic graphene).
The number of electrons (per area and per time) perpendicularly crossing the graphene plane  with total energy between $E$ and $E+dE$ and normal energy between $E_x$ and $E_x+dE_x$ is given by

\begin{eqnarray}
N(E,E_x)dEdE_{x}\equiv \frac{2f_{F-D}(E)}{(2\pi)^2}\int\int\int_{E,E_x}v_{x}d^2kdk_x \\ \nonumber
=\frac{1}{\pi\hbar^{3}v_f^{2}}E_{p}f_{FD}(E)dE_{p}dE_x.\label{3}
\end{eqnarray}

In the derivation of Eq. (4), we have assumed that the normal energy component of electrons is $E_x=\hbar^{2}k_{x}^{2}/2m$.
This assumption may be justified by the reasoning that graphene has atomic thickness for which the electrons are confined in a finite quantum well in the normal direction.
The barrier height of the quantum well is assumed to be the intrinsic work function of a single layer graphene, which may be different from bulk graphite.
Due to the quantum confinement, we solve the time-independent Schr$\ddot{o}$dinger equation of electrons traveling in the normal direction to obtain the energy of the ground state, which is  $E_x$ as given in the equations above.
Using Eq. (4), we calculate the total number of electrons with normal energy between $E_x$ and $E_x+dE_x$ at equilibrium condition, which gives
\begin{equation}
N(E_x)dE_{x}=\frac{dE_x}{\pi\hbar^{3}v_f^{2}}\int_{E_x}^{\infty}(E-E_x)f_{F-D}(E)dE.\label{4}
\end{equation}

To solve Eq. (\ref{4}) analytically, we make the follow assumptions, which will be verified by comparing the analytical result with numerical calculation as shown in Fig. 2(b).
Firstly, we assume that only the electrons with energy greater than or equal to the work function of cathode (graphene) can be emitted, which means electron tunneling is completely omitted.
This is reasonable as it is the definition of thermionic electron emission.
It also  implies that only high-energy component of the Fermi-Dirac distribution is important, for which we may replace the Fermi-Dirac distribution with the Maxwell-Boltzmann distribution $f_{M-B}(E)=\mbox{exp}[-(E-E_{F})/k_{B}T]$.
In doing so,  Eq.(\ref{4}) is simplified to become
\begin{eqnarray}
N(E_x)dE_{x}=\frac{k_{B}^{2}T^{2}}{\pi\hbar^{3}v_f^{2}}exp[-\frac{E_{x}-E_{F}}{k_{B}T}]dE_x.
\end{eqnarray}

For thermionic emission, the current density of the emitted electrons along the direction perpendicular to graphene plane is calculated by
\begin{equation}
J(E_{F},T)=\int_{\Phi}^{\infty}eN(E_x)dE_{x}. \label{45}
\end{equation}
In solving Eqs. (6) and (7), we obtain an analytical formula for thermionic electron emission from a single-layer graphene excluding the effect of substrate, which is
\begin{equation}
J(E_{F},T)=\frac{ek_{B}^{3}T^{3}}{\pi\hbar^{3}v_f^{2}}exp[-\frac{\Phi-E_{F}}{k_{B}T}]\label{55}.
\end{equation}
It is clear that Eq. (8) is independent of mass $m$, solving the inconsistency of RD law having a finite mass term in the prefactor $A$, which is troublesome in applying RD law for electron emission from a single layer graphene. Note the work function is considered to be temperature-independent in our paper. Based on this interesting finding and the importance of electron band structure determining the amount of electron thermionic emission, we speculate that all the materials with linear band structure may have the same scaling law reported here.

It is common that an external and small voltage is used to collect thermionically emitted electrons at the anode.
With the applied voltage, the potential barrier will be reduced to $\Phi_{eff}$ = $\Phi - \Delta \Phi$ (due to Schottky effect) as shown in the Fig. 1b (dark blue line).
Under this situation, the equation is modified as
\begin{equation}
J(E_{F},T)= \beta T^{3} exp[-\frac{\Phi-E_{F}-\Delta\Phi}{k_{B}T}]\label{7},
\end{equation}
where $\beta=ek_{B}^{3}/\pi \hbar^3 v_{f}^{2}=115.8$ Am$^{2}$K$^{-3}$, $\Delta\Phi=(e^{3}F/4\pi\epsilon_o)^{1/2}$ represents the reduction in the potential barrier, and $F$ ($<$ 0.1 V/nm) is the electric field \cite{MurphGood1956}.
At higher field $F >$ 1 V/nm,  quantum tunneling begins to dominate over the Schottky emission process, which requires new model \cite{xianlongwei2013}, and it will not be addressed in this paper.
Throughout this paper, we have assumed $F$ = 0 unless it is specified.

To verify Eq.(\ref{55}), it is important to have a more realistic surface potential profile $V(x)$ in order to estimate the contribution of  electron tunneling through the potential barrier.
In doing so, the classical image charge potential $V_{im}$ is arbitrarily included in $V(x)$ (near to the cathode), and $V_{im}$  is \cite{Smythe}
\begin{equation}
V_{im}(x)=-\frac{e^2}{4\pi\epsilon_o}[\frac{1}{2x}+\sum^{\infty}_{n=1}\{\frac{nD}{(nD^2)-x^2}-\frac{1}{D}\}],\label{11}
\end{equation}
which can be simplified to be  \cite{Simmons1963}
\begin{equation}
V_{im}(x)=-\frac{1.15\alpha D^2}{x(x-D)}.
\end{equation}
Here, $\alpha=e^2\ln2/8\pi\epsilon_{0} D$, $D$ is the gap spacing, and $\epsilon_o$ is the vacuum permittivity.
The potential barrier $V(x)$ in between the gap is
\begin{equation}
V(x)=\Phi- V_{im}(x)= \Phi-\frac{1.15\alpha D^2}{x(x-D)}.
\end{equation}
Note that the singularities at $x$ = 0 and $x = D$ can be avoided by restricting the positive values of $x$ (within $ 0 < x < D$) to have a positive $V(x)$.
This simplified image charge model had been verified by using a quantum image charge model \cite{Koh2008}, and it is valid for $D >$ 10 nm.

To determine the effect of electron tunneling through the barrier $V(x)$, we calculate the tunneling probability $D(E_{x})$ by solving
\begin{eqnarray}
D(E_{x})&=& exp[-\frac{4\pi\sqrt{2m}}{h}\int^{x_{2}}_{x_{1}}[(V(x)-E_{x})]^{1/2}dx]\\\nonumber
&\approx & exp[-\tilde{B}(H-E_{x})^{1/2}],\label{12}
\end{eqnarray}
where $H= \Phi-1.15\alpha/b\ln[\frac{(1+b)^2}{(1-b)^2}]$ is the modified barrier height, $b=\sqrt{1-4.6\alpha/\Phi}$ and $\tilde{B}=2(x_{2}-x_{1})(2m)^{1/2}/h$.
The values of $x_{1}$ and $x_{2}$ are calculated by solving $V(x)=0$.
The amount of the tunneling current density is calculated by
\begin{eqnarray}
J_{T}&=&e\int^{\infty}_{0}N(E_{x})D(E_{x})dE_{x}\\\nonumber \label{133}
 &\approx&\frac{e^{-\tilde{B}H^{1/2}}}{\pi\hbar^{3}v_{f}^{2}/e}(\frac{1}{3}E_{f}^{3}-2(k_{B}T)^{3}  \mbox{polylog}[3,-e^{-\frac{E_{f}}{k_{B}T}}]),
\end{eqnarray}
where polylog[3,x] is the polylogarithm function \cite{Wooddc1992}.

In the derivation of Eq. (14), we have assumed that only electrons around the Fermi level participate in the tunneling process.
Based on the default parameters used in our calculation: $E_f = $ 0.083 eV, $\Phi$ = 4.514 eV and $v_f = 10^6$ m/s, the term $e^{-\tilde{B}H^{1/2}}$ is nearly zero, which implies that
the amount of the tunneling current density $J_{T}$ can be ignored as compared with the thermionic current density given by Eq. (\ref{55}).

\section{Thermionic emission from Graphene}

In Fig. 2(a), the thermionic emitted current density $J$ based on Eq. (\ref{55}) is plotted as a function of temperature $T$ (log-log scale) for various  $\Phi$ = 3, 4, and 5 eV [solid lines] at fixed $E_f = $ 0.083 eV, and $v_f = 10^6$ m/s.
The results based on the RD law are also presented [dashed lines] for comparison. Here, we have assumed that the work function does not change with the temperature.
For a given work function, a new scaling of $T^{3}$ is predicted by our model, which is different from the traditional RD scaling of $T^2$.
In the inset,  we plot $ln(J/T^3)$ (left $y$ axis for our model) and $ln(J/T^2)$ (right $y$ axis for RD law) versus $1/T$ for work functions = 3 to 5 eV.
%Compared with the RD law, our model gives a steeper slope for same work function, which indicates that
%That is to say, our model can be fitted to given experiment data \cite{jiangkaili2014} with a lower work function.

In Fig. 2(b), we compare the analytical formula Eq. (\ref{55}) [red line] with the numerical solution of Eq. (\ref{45}) [circles] up to $T$ = 2500 K (at  $E_f$ = 0.083 eV and $\Phi$ = 4.514 eV).
The comparison shows a good agreement up to $T$ = 2000 K.
The deviation at higher temperature is due to the approximation of Maxwell-Boltzmann distribution to Fermi-Dirac distribution in solving the integral.
This comparison implies that Eq. (\ref{55}) is accurate up to 2000 K (a practical temperature).

In Fig. 3(a), the thermionic current density $J$ [from Eq. (\ref{55})] is calculated as a function of $ T$ (up to 2000 K) for various Fermi energy level $E_f$ = 0.083 to 0.4 eV.
From the figure, we see that increasing $E_f$ can drastically enhance $J$.
In order to decrease the temperature to $T$ = 900 to 1000 K with $J >$ 10 A/m$^2$, it is required to have a higher $E_f$ = 0.8 to 0.9 eV and also a lower work function $\Phi$ = 2.5 or 3 eV as shown in Fig. 3b.
Note that the calculated $J$  is sensitive at around $E_f$ = 0.8 and 0.9 eV and $\Phi$ = 2.5 to 3 eV.
For example,  if we use lower $E_f$ = 0.8 eV and higher $\Phi$ = 3 eV, $J$ is significantly to $J <$ 1 A/m$^2$  (see the insert in Fig. 3b).
It is worth to note that Fermi level can be tuned experimentally over a wide range (0.5 to 0.85 eV) via different methods such as chemical doping \cite{Giovannetti2008}, and electrostatic gate voltage \cite{Englund2013}.
Some recent studies has also suggested that graphene can be engineered to have a lower work function (2.5 to 3 eV) \cite{Kralj2014}.

There is a recent experiment measuring the thermionic electron emission from a single-layer graphene (suspended on a CNT film) in the range of $T$ = 1600 to 1750 K [see Fig. 4 in Ref. \cite{jiangkaili2014}].
In Fig. 4(a), we fit the experimental data [black square symbol] to our model with a work function of about $\Phi$ = 4.514 eV and $E_F$ = 0.083 eV, which shows a better agreement (black solid line) as compared with using traditional RD law with 4.66 eV for the bulk graphite (dashed blue line).
In comparison, the RD law at 4.514 eV (dashed red line) shows much larger values and our model at 4.66 eV predicts much lower current values than the experimental data.
This extracted work function (4.514 eV) is in excellent agreement with the experimental measured value around 4.5 eV \cite{kunxu2012nl}, and also the theoretical value by DFT \cite{Koch2014JPCC}.
Theoretically, the reduction of work function is expected because when the thickness of bulk graphite is reduced down to single-layer graphene, the quantum size effect in the normal direction becomes very important, then the size-reduce-induced confinement causes the upward shift in electron total energy, and thus the work function of graphene become smaller [see Section 4.1 in the Ref.\cite{changqing2011}].

On the other hand, if the graphene is placed on different metallic substrates, another recent experiment \cite{Starodub2012} has shown that the data is well described by the traditional RD law, as shown in the Fig.4(b).
Here, the type of substrate will affect the amount of current significantly with a fitted and effective work function (based on RD law) ranging from 3.3 to 4.6 eV including effects of the substrate \cite{Starodub2012}.
The findings [reported in Figs. 4(a) and 4(b)] indicate that the metallic based substrate will play an important role for thermionic emission from graphene. Note that the effects of the substrate have been completely ignored in our model, which is beyond the scope of this paper.

\section{Thermionic energy convertor (TIC) }

By ignoring other energy losses, we calculate the ideal efficiency $\eta$ of a thermionic energy convertor (TIC) in using graphene as cathode (4.514 eV) at cathode temperature $T_c$, and metallic material as anode of different work function $\Phi_a$ and temperature $T_a$.
The value of $\eta$ is calculated by
\begin{equation}
   \eta=\frac{(J_{c}-J_{a})(\Phi_{c}-\Phi_{a})}{J_{c}(\Phi_{c}+2k_{B}T_{c})-J_{a}(\Phi_{a}+2k_{B}T_{a})}.\label{6}
 \end{equation}
%where $T_a$ and $T_c$ denote the anode temperature and cathode temperature, respectively; $\Phi_{a}$ and $\Phi_{c}$ represent the work function of anode and of cathode respectively; $J_a$ and $J_c$ is, respectively, the current density from anode to %cathode, and from the cathode to anode.
Here, we have used Eq. (\ref{55}) to calculate $J_c$, and the RD law for $J_a$.

Figure 5 shows the calculated $\eta$ (solid lines) as a function $\Phi_a$ = 0.5 to 4.5 eV for different $T_a$ [K] = 300, 500, 800 and 1000 for a fixed graphene cathode property: $T_c$ = 1800 K, $\Phi_c$ = 4.514 eV and $E_f$ = 0.083 eV.
From the figure, we see that, for a given $T_a$, there is an optimal value of anode work function $\Phi_a$ for which the TIC will have a maximal value of $\eta_{max}$.
As an example,  we have $\eta_{max} \approx$ 0.45 at $T_a$ = 800 K, $T_c$ = 1800 K, and $\Phi_a \approx$ 2.25 eV (blue line).
This behavior remains valid for $T_c$  down to 900 K [see the inset].
The maximum efficiency $\eta_{max}$ (left $y$ axis) and its corresponding optimal values of anode work function $\Phi_a$ (right $y$ axis) are plotted in the inset as a function of anode temperature $T_a$ = 300 to 800 K for two cathode temperature $T_c$ = 900 K (black dashed lines) and 1800 K (red solid lines).
Using this insert, we can immediately know the maximum efficiency $\eta_{max}$ of TIC and obtain the required values of $T_a$ and $\Phi_a$.
At low $T_c$ = 900 K (see black dashed lines in the inset), we will need $T_{a} \approx$ 425 K and $\Phi_{a} \approx$ 2.5 eV to have $\eta_{max} \approx$ 0.45.
Note that the most recent work shows that LaB$_6$  heterostructure may have work function around to 2 to 3 eV \cite{voss2014}, and it may be a suitable candidate for the anode material for the proposed high efficiency TIC shown in Fig. 5.

%\begin{equation}
%\eta_{max}=-0.2\Phi_{a}+0.933.\label{17}
%\end{equation}

\section{Conclusions}

In summary, we have proposed to use single-layer graphene as the cathode material in the design of a high efficiency thermionic energy convertor (TIC).
In order to provide some quantitative design parameters, we first develop a model to calculate the electron thermionic emission from a single-layer suspended graphene by ignoring the effect of the substrate.
From our model, an analytical formula [Eq. (\ref{55})] is formulated, which has been verified with a numerical calculation and compared with experimental results.
Our findings suggest that the traditional thermionic emission law governed by the well-known Richardson-Dunshman (RD) equation is no longer valid if the effect of the substrate is not important.
Our model predicts a new scaling of $T^{3}$, which is different from the classical RD scaling of $T^{2}$.

By considering that the graphene can be engineered to have a higher Fermi energy (around 0.8 to 0.9 eV) \cite{Giovannetti2008,Englund2013}, and to have a lower work function (to 2.5 to 3 eV) \cite{Kralj2014},
it is possible to have a graphene-based cathode to emit a sufficiently high current density ($>$ 10 A/m$^2$) at $T$ = 700 to 1000 K.

A proposed TIC with a single layer graphene as cathode (work function = 4.514 eV) and metallic electrode as anode of different work function is designed for which the optimal anode work function and anode temperature are calculated to obtain a maximal efficiency.
It is possible to have a TIC at an efficiency of about 45 percent with a cathode temperature of 900 K and with a metallic anode of about 2.5 eV work function (like LaB$_6$ anode) at around 425 K.
These findings may shed new light on developing thermionic cathode using graphene and also thermal energy harvesting device like TIC.
The work is also useful in the development of novel materials with low work function, like the new LaB$_6$ heterostructure with a lowering work function of 0.46 eV \cite{voss2014}.

\section*{Acknowledgments}
Authors would like to thank Dr. Peng Liu and Prof. Kaili Jiang for helpful discussion and kindly providing us with their original experimental data. In addition, we are also grateful for referee's constructive comments on the manuscript. This work was supported by SUTD (SRG EPD 2011 014) and SUTD-MIT IDC Grant (IDG21200106 and IDD21200103).
LKA acknowledge the support of AFOAR AOARD grant (14-2110).
%\bibliographystyle{prsty}

%\bibliographystyle{nature}
%\bibliography{liang-v2}

\begin{thebibliography}{38}
\expandafter\ifx\csname natexlab\endcsname\relax\def\natexlab#1{#1}\fi
\expandafter\ifx\csname bibnamefont\endcsname\relax
  \def\bibnamefont#1{#1}\fi
\expandafter\ifx\csname bibfnamefont\endcsname\relax
  \def\bibfnamefont#1{#1}\fi
\expandafter\ifx\csname citenamefont\endcsname\relax
  \def\citenamefont#1{#1}\fi
\expandafter\ifx\csname url\endcsname\relax
  \def\url#1{\texttt{#1}}\fi
\expandafter\ifx\csname urlprefix\endcsname\relax\def\urlprefix{URL }\fi
\providecommand{\bibinfo}[2]{#2}
\providecommand{\eprint}[2][]{\url{#2}}

\bibitem[{\citenamefont{Richardson}(2003)}]{Richardson2003}
\bibinfo{author}{\bibfnamefont{O.~W.} \bibnamefont{Richardson}},
  \emph{\bibinfo{title}{Thermionic Emission from Hot Bodies,}}
  (\bibinfo{publisher}{Wexford College Press}, \bibinfo{address}{London, UK},
  \bibinfo{year}{2003}).

\bibitem[{\citenamefont{Terence D.~Musho and Walker}(2013)}]{walker2013}
\bibinfo{author}{\bibfnamefont{J.~L.~D.} \bibnamefont{Terence D.~Musho},
  \bibfnamefont{William F.~Paxton}} \bibnamefont{and}
  \bibinfo{author}{\bibfnamefont{D.~G.} \bibnamefont{Walker}},
  \bibinfo{title}{Quantum Simulation of Thermionic Emission from Diamond Films,}
  \bibinfo{journal}{J. Vac. Technol. B} \textbf{\bibinfo{volume}{31}},
  \bibinfo{pages}{021401} (\bibinfo{year}{2013}).

\bibitem[{\citenamefont{Novoselov et~al.}(2004)\citenamefont{Novoselov, Geim,
  Morozov, Jiang, Zhang, Dubonos, Grigorieva, and Firsov}}]{Novoselov2004}
\bibinfo{author}{\bibfnamefont{K.~S.} \bibnamefont{Novoselov}},
  \bibinfo{author}{\bibfnamefont{A.~K.} \bibnamefont{Geim}},
  \bibinfo{author}{\bibfnamefont{S.~V.} \bibnamefont{Morozov}},
  \bibinfo{author}{\bibfnamefont{D.}~\bibnamefont{Jiang}},
  \bibinfo{author}{\bibfnamefont{Y.}~\bibnamefont{Zhang}},
  \bibinfo{author}{\bibfnamefont{S.~V.} \bibnamefont{Dubonos}},
  \bibinfo{author}{\bibfnamefont{I.~V.} \bibnamefont{Grigorieva}},
  \bibnamefont{and} \bibinfo{author}{\bibfnamefont{A.~A.} \bibnamefont{Firsov}},
  \bibinfo{title}{Effect in Atomically Thin Carbon Films,}
  \bibinfo{journal}{Nature}
  \textbf{\bibinfo{volume}{306}}, \bibinfo{pages}{666} (\bibinfo{year}{2004}).

\bibitem[{\citenamefont{Novoselov et~al.}(2005)\citenamefont{Novoselov, Geim,
  Morozov, Jiang, Katsnelson, Grigorieva, Dubonos, and Firsov}}]{Novoselov2005}
\bibinfo{author}{\bibfnamefont{K.~S.} \bibnamefont{Novoselov}},
  \bibinfo{author}{\bibfnamefont{A.~K.} \bibnamefont{Geim}},
  \bibinfo{author}{\bibfnamefont{S.~V.} \bibnamefont{Morozov}},
  \bibinfo{author}{\bibfnamefont{D.}~\bibnamefont{Jiang}},
  \bibinfo{author}{\bibfnamefont{M.~I.} \bibnamefont{Katsnelson}},
  \bibinfo{author}{\bibfnamefont{I.~V.} \bibnamefont{Grigorieva}},
  \bibinfo{author}{\bibfnamefont{S.~V.} \bibnamefont{Dubonos}},
  \bibnamefont{and} \bibinfo{author}{\bibfnamefont{A.~A.}
  \bibnamefont{Firsov}},
  \bibinfo{title}{Two-dimensional gas of massless Dirac fermions in graphene,}
  \bibinfo{journal}{Nature}
  \textbf{\bibinfo{volume}{438}}, \bibinfo{pages}{197} (\bibinfo{year}{2005}).

\bibitem[{\citenamefont{Balandin et~al.}(2008)\citenamefont{Balandin, Ghosh,
  Bao, Calizo, Teweldebrhan, Miao, and Lau}}]{Balandin2008}
\bibinfo{author}{\bibfnamefont{A.~A.} \bibnamefont{Balandin}},
  \bibinfo{author}{\bibfnamefont{S.}~\bibnamefont{Ghosh}},
  \bibinfo{author}{\bibfnamefont{W.}~\bibnamefont{Bao}},
  \bibinfo{author}{\bibfnamefont{I.}~\bibnamefont{Calizo}},
  \bibinfo{author}{\bibfnamefont{D.}~\bibnamefont{Teweldebrhan}},
  \bibinfo{author}{\bibfnamefont{F.}~\bibnamefont{Miao}}, \bibnamefont{and}
  \bibinfo{author}{\bibfnamefont{C.~N.} \bibnamefont{Lau}},
  \bibinfo{title}{Superior thermal conductivity of single-layer graphene,}
  \bibinfo{journal}{Nano Letters} \textbf{\bibinfo{volume}{8}},
  \bibinfo{pages}{902} (\bibinfo{year}{2008}).

\bibitem[{\citenamefont{Sun et~al.}(2011)\citenamefont{Sun, Ang, Shiffler, and
  Luginsland}}]{Sun2011}
\bibinfo{author}{\bibfnamefont{S.}~\bibnamefont{Sun}},
  \bibinfo{author}{\bibfnamefont{L.~K.} \bibnamefont{Ang}},
  \bibinfo{author}{\bibfnamefont{D.}~\bibnamefont{Shiffler}}, \bibnamefont{and}
  \bibinfo{author}{\bibfnamefont{J.~W.} \bibnamefont{Luginsland}},
  \bibinfo{title}{Klein tunneling model of low energy electron field emission from single-layer graphene sheet,}
  \bibinfo{journal}{Appl. Phys. Lett.} \textbf{\bibinfo{volume}{99}},
  \bibinfo{pages}{013112} (\bibinfo{year}{2011}).

\bibitem[{\citenamefont{Liang et~al.}(2013)\citenamefont{Liang, Sun, and
  Ang}}]{shijuna}
\bibinfo{author}{\bibfnamefont{S.-J.} \bibnamefont{Liang}},
  \bibinfo{author}{\bibfnamefont{S.}~\bibnamefont{Sun}}, \bibnamefont{and}
  \bibinfo{author}{\bibfnamefont{L.~K.} \bibnamefont{Ang}},
  \bibinfo{title}{Over-barrier side-band electron emission from graphene with a time-oscillating potential,}
  \bibinfo{journal}{Carbon} \textbf{\bibinfo{volume}{61}}, \bibinfo{pages}{294}
  (\bibinfo{year}{2013}).

\bibitem[{\citenamefont{Liang and Ang}(2014)}]{SJLiang2014}
\bibinfo{author}{\bibfnamefont{S.-J.} \bibnamefont{Liang}} \bibnamefont{and}
  \bibinfo{author}{\bibfnamefont{L.~K.} \bibnamefont{Ang}},
  \bibinfo{title}{Chiral Tunneling-Assisted Over-Barrier Electron. Emission from Graphene,}
  \bibinfo{journal}{IEEE Trans. Electron Devices}
  \textbf{\bibinfo{volume}{61}}, \bibinfo{pages}{1764} (\bibinfo{year}{2014}).

\bibitem[{\citenamefont{Walt A.~de Heer and Ugarte}(1995)}]{WaltAdeHeer1995}
\bibinfo{author}{\bibfnamefont{A.~C.} \bibnamefont{Walt A.~de Heer}}
  \bibnamefont{and} \bibinfo{author}{\bibfnamefont{D.}~\bibnamefont{Ugarte}},
  \bibinfo{title}{A Carbon Nanotube Field-Emission Electron Source,}
  \bibinfo{journal}{Science} \textbf{\bibinfo{volume}{270}},
  \bibinfo{pages}{1179} (\bibinfo{year}{1995}).

\bibitem[{\citenamefont{Jean-Marc~Bonard
  et~al.}(2002)\citenamefont{Jean-Marc~Bonard, Noury, and
  Weiss}}]{JeanMarcBonard2002}
\bibinfo{author}{\bibfnamefont{C.~K. R.~K.} \bibnamefont{Jean-Marc~Bonard},
  \bibfnamefont{Mirko~Croci}},
  \bibinfo{author}{\bibfnamefont{O.}~\bibnamefont{Noury}}, \bibnamefont{and}
  \bibinfo{author}{\bibfnamefont{N.}~\bibnamefont{Weiss}},
  \bibinfo{title}{Carbon nanotube films as electron field emitters,}
  \bibinfo{journal}{Carbon} \textbf{\bibinfo{volume}{40}},
  \bibinfo{pages}{1715} (\bibinfo{year}{2002}).

\bibitem[{\citenamefont{Sarker and Khondaker}(2012)}]{Khondaker2012}
\bibinfo{author}{\bibfnamefont{B.~K.} \bibnamefont{Sarker}} \bibnamefont{and}
  \bibinfo{author}{\bibfnamefont{S.~I.} \bibnamefont{Khondaker}},
  \bibinfo{title}{Thermionic emission and tunneling at carbon nanotube-organic semiconductor interface,}
  \bibinfo{journal}{ACS Nano} \textbf{\bibinfo{volume}{6}},
  \bibinfo{pages}{4993} (\bibinfo{year}{2012}).

\bibitem[{\citenamefont{Cheng and Zhou}(2003)}]{Otto2003}
\bibinfo{author}{\bibfnamefont{Y.}~\bibnamefont{Cheng}} \bibnamefont{and}
  \bibinfo{author}{\bibfnamefont{O.}~\bibnamefont{Zhou}},
  \bibinfo{title}{Electron field emission from carbon nanotubes,}
   \bibinfo{journal}{C.
  R. Physique} \textbf{\bibinfo{volume}{4}}, \bibinfo{pages}{1021}
  (\bibinfo{year}{2003}).

\bibitem[{\citenamefont{Shi-Dong~Liang and Xu}(2006)}]{liangshidong2006}
\bibinfo{author}{\bibfnamefont{S.~Z.~D.} \bibnamefont{Shi-Dong~Liang},
  \bibfnamefont{N.Y.~Huang}} \bibnamefont{and}
  \bibinfo{author}{\bibfnamefont{N.}~\bibnamefont{Xu}},
  \bibinfo{title}{Quantum effects in the field emission of carbon nanotubes,}
   \bibinfo{journal}{J. Vac. Technol. B} \textbf{\bibinfo{volume}{B24}}, \bibinfo{pages}{983}
  (\bibinfo{year}{2006}).

\bibitem[{\citenamefont{Liang and Chen}(2008)}]{liangshidong2008}
\bibinfo{author}{\bibfnamefont{S.-D.} \bibnamefont{Liang}} \bibnamefont{and}
  \bibinfo{author}{\bibfnamefont{L.}~\bibnamefont{Chen}},
  \bibinfo{title}{Generalized Fowler-Nordheim Theory of Field Emission of Carbon Nanotubes,}
  \bibinfo{journal}{Phys. Rev. Lett.} \textbf{\bibinfo{volume}{101}},
  \bibinfo{pages}{027602} (\bibinfo{year}{2008}).

\bibitem[{\citenamefont{Liang and Chen}(2010)}]{liangshidong2010}
\bibinfo{author}{\bibfnamefont{S.-D.} \bibnamefont{Liang}} \bibnamefont{and}
  \bibinfo{author}{\bibfnamefont{L.}~\bibnamefont{Chen}},
  \bibinfo{title}{Theories of field and thermionic electron emissions from carbon nanotubes,}
  \bibinfo{journal}{J.
  Vac. Technol. B} \textbf{\bibinfo{volume}{B28}}, \bibinfo{pages}{C2A50}
  (\bibinfo{year}{2010}).

\bibitem[{\citenamefont{X~Wei}(2014)}]{xianlongwei2014}
\bibinfo{author}{\bibfnamefont{Q.~C. L.~P.} \bibnamefont{X~Wei},
  \bibfnamefont{S~Wang}},
  \bibinfo{title}{Breakdown of Richardson's law in electron emission from individual self-Joule-heated carbon nanotubes,}
  \bibinfo{journal}{Sci. Rep.}
  \textbf{\bibinfo{volume}{4}}, \bibinfo{pages}{5102} (\bibinfo{year}{2014}).

\bibitem[{\citenamefont{X~Wei}(2011)}]{xianlongwei2011}
\bibinfo{author}{\bibfnamefont{Q.~C. Y. B. L.~P.} \bibnamefont{X~Wei},
  \bibfnamefont{D~Golberg}},
  \bibinfo{title}{Phonon-Assisted Electron Emission from Individual Carbon Nanotubes,}
  \bibinfo{journal}{Nano letters}
  \textbf{\bibinfo{volume}{11}}, \bibinfo{pages}{734} (\bibinfo{year}{2011}).

\bibitem[{\citenamefont{Yoon et~al.}(2014)\citenamefont{Yoon, Forsythe,
  L.~Wang, Watanabe, Taniguchi, Hone, Kim, and Hami}}]{Yoon2014}
\bibinfo{author}{\bibfnamefont{H.}~\bibnamefont{Yoon}},
  \bibinfo{author}{\bibfnamefont{C.}~\bibnamefont{Forsythe}},
  \bibinfo{author}{\bibfnamefont{N.~T.} \bibnamefont{L.~Wang}},
  \bibinfo{author}{\bibfnamefont{K.}~\bibnamefont{Watanabe}},
  \bibinfo{author}{\bibfnamefont{T.}~\bibnamefont{Taniguchi}},
  \bibinfo{author}{\bibfnamefont{J.}~\bibnamefont{Hone}},
  \bibinfo{author}{\bibfnamefont{P.}~\bibnamefont{Kim}}, \bibnamefont{and}
  \bibinfo{author}{\bibfnamefont{D.}~\bibnamefont{Hami}},
  \bibinfo{title}{Measurement of collective dynamical mass of Dirac fermions in graphene,}
  \bibinfo{journal}{Nature Nanotechnology} \textbf{\bibinfo{volume}{9}},
  \bibinfo{pages}{594} (\bibinfo{year}{2014}).

\bibitem[{\citenamefont{M.~Zebarjadi and Chen}(2011)}]{Zebarjadi2012}
\bibinfo{author}{\bibfnamefont{M.~D. Z.~R.} \bibnamefont{M.~Zebarjadi},
  \bibfnamefont{K.~Esfarjani}} \bibnamefont{and}
  \bibinfo{author}{\bibfnamefont{G.}~\bibnamefont{Chen}},
  \bibinfo{title}{Perspectives on thermoelectrics: from fundamentals to device applications,}
  \bibinfo{journal}{Energy and Environmental Science}
  \textbf{\bibinfo{volume}{5}}, \bibinfo{pages}{5147} (\bibinfo{year}{2011}).

\bibitem[{\citenamefont{Hatsopoulos and Gyftopoulos}(1974)}]{Hatsopoulos1974}
\bibinfo{author}{\bibfnamefont{G.~N.} \bibnamefont{Hatsopoulos}}
  \bibnamefont{and} \bibinfo{author}{\bibfnamefont{E.~P.}
  \bibnamefont{Gyftopoulos}}, \emph{\bibinfo{title}{Thermionic Energy
  Conversion I,}} (\bibinfo{publisher}{MIT Press}, \bibinfo{address}{Cambridge,
  MA}, \bibinfo{year}{1974}).

\bibitem[{\citenamefont{S.~Meir and Mannhart}(2013)}]{Meir2013}
\bibinfo{author}{\bibfnamefont{T.~H.~G.} \bibnamefont{S.~Meir},
  \bibfnamefont{C.~Stphanos}} \bibnamefont{and}
  \bibinfo{author}{\bibfnamefont{J.}~\bibnamefont{Mannhart}},
  \bibinfo{title}{Highly-efficient thermoelectronic conversion of solar energy and heat into electric power,}
  \bibinfo{journal}{Journal of Renewable and Sustainable Energy}
  \textbf{\bibinfo{volume}{5}}, \bibinfo{pages}{043127} (\bibinfo{year}{2013}).

\bibitem[{\citenamefont{Schwede et~al.}(2013)\citenamefont{Schwede, Sarmiento,
  Narasimhan, Rosenthal, Schmitt, Bargatin, Sahasrabuddhe, Howe, Harris, Shen
  et~al.}}]{Jared2013}
\bibinfo{author}{\bibfnamefont{J.~W.} \bibnamefont{Schwede}},
  \bibinfo{author}{\bibfnamefont{T.}~\bibnamefont{Sarmiento}},
  \bibinfo{author}{\bibfnamefont{V.~K.} \bibnamefont{Narasimhan}},
  \bibinfo{author}{\bibfnamefont{S.~J.} \bibnamefont{Rosenthal}},
  \bibinfo{author}{\bibfnamefont{F.}~\bibnamefont{Schmitt}},
  \bibinfo{author}{\bibfnamefont{I.}~\bibnamefont{Bargatin}},
  \bibinfo{author}{\bibfnamefont{K.}~\bibnamefont{Sahasrabuddhe}},
  \bibinfo{author}{\bibfnamefont{R.~T.} \bibnamefont{Howe}},
  \bibinfo{author}{\bibfnamefont{J.~S.} \bibnamefont{Harris}},
  \bibinfo{author}{\bibfnamefont{Z.~X.} \bibnamefont{Shen}},
  \bibnamefont{et~al.},
  \bibinfo{title}{Photon-enhanced thermionic emission from heterostructures with low interface recombination,}
  \bibinfo{journal}{Nature communications}
  \textbf{\bibinfo{volume}{4}}, \bibinfo{pages}{1576} (\bibinfo{year}{2013}).

\bibitem[{\citenamefont{Schwede et~al.}(2010)\citenamefont{Schwede, Bargatin,
  Riley, Hardin, Rosenthal, Su, Schmitt, Pianetta, Howe, Shen
  et~al.}}]{Jared2010}
\bibinfo{author}{\bibfnamefont{J.~W.} \bibnamefont{Schwede}},
  \bibinfo{author}{\bibfnamefont{L.}~\bibnamefont{Bargatin}},
  \bibinfo{author}{\bibfnamefont{D.~C.} \bibnamefont{Riley}},
  \bibinfo{author}{\bibfnamefont{B.~E.} \bibnamefont{Hardin}},
  \bibinfo{author}{\bibfnamefont{S.~J.} \bibnamefont{Rosenthal}},
  \bibinfo{author}{\bibfnamefont{Y.}~\bibnamefont{Su}},
  \bibinfo{author}{\bibfnamefont{F.}~\bibnamefont{Schmitt}},
  \bibinfo{author}{\bibfnamefont{P.}~\bibnamefont{Pianetta}},
  \bibinfo{author}{\bibfnamefont{R.~T.} \bibnamefont{Howe}},
  \bibinfo{author}{\bibfnamefont{Z.-X.} \bibnamefont{Shen}},
  \bibnamefont{et~al.},
  \bibinfo{title}{Photon enhanced thermionic emission for solar concentrator systems,}
   \bibinfo{journal}{Nature Materials}
  \textbf{\bibinfo{volume}{9}}, \bibinfo{pages}{726} (\bibinfo{year}{2010}).

\bibitem[{\citenamefont{Gideon~Segev and Kribus}(2013)}]{Segev2013}
\bibinfo{author}{\bibfnamefont{Y.~R.} \bibnamefont{Gideon~Segev}}
  \bibnamefont{and} \bibinfo{author}{\bibfnamefont{A.}~\bibnamefont{Kribus}},
  \bibinfo{title}{Loss mechanisms and back surface field effect in photon enhanced thermionic emission converters,}
  \bibinfo{journal}{J. App. Phys.} \textbf{\bibinfo{volume}{114}},
  \bibinfo{pages}{044505} (\bibinfo{year}{2013}).

\bibitem[{\citenamefont{E.~Starodub and McCarty}(2012)}]{Starodub2012}
\bibinfo{author}{\bibfnamefont{N.~C.~B.} \bibnamefont{E.~Starodub}}
  \bibnamefont{and} \bibinfo{author}{\bibfnamefont{K.~F.}
  \bibnamefont{McCarty}},
  \bibinfo{title}{Viable thermionic emission from graphene-covered metals,}
   \bibinfo{journal}{App. Phys. Lett.}
  \textbf{\bibinfo{volume}{100}}, \bibinfo{pages}{181604}
  (\bibinfo{year}{2012}).

\bibitem[{\citenamefont{Feng~Zhu and Fan}(2014)}]{jiangkaili2014}
\bibinfo{author}{\bibfnamefont{P.~L. K. J. Y. W. Y. W. J.~W.}
  \bibnamefont{Feng~Zhu}, \bibfnamefont{Xiaoyang~Lin}} \bibnamefont{and}
  \bibinfo{author}{\bibfnamefont{S.}~\bibnamefont{Fan}},
  \bibinfo{title}{Heating graphene to incandescence and the measure-ment of its work function by the thermionic emission method,}
   \bibinfo{journal}{Nano
  Research} \textbf{\bibinfo{volume}{7}}, \bibinfo{pages}{553}
  (\bibinfo{year}{2014}).

\bibitem[{\citenamefont{Wallace}(1974)}]{Wallace1974}
\bibinfo{author}{\bibfnamefont{P.~R.} \bibnamefont{Wallace}},
\bibinfo{title}{The Band Theory of Graphite,}
  \bibinfo{journal}{Phys. Rev.} \textbf{\bibinfo{volume}{71}},
  \bibinfo{pages}{622} (\bibinfo{year}{1974}).

\bibitem[{\citenamefont{Neto et~al.}(2009)\citenamefont{Neto, Guinea, Peres,
  Novoselov, and Geim}}]{Castroneto2009}
\bibinfo{author}{\bibfnamefont{A.~H.~C.} \bibnamefont{Neto}},
  \bibinfo{author}{\bibfnamefont{F.}~\bibnamefont{Guinea}},
  \bibinfo{author}{\bibfnamefont{N.~M.~R.} \bibnamefont{Peres}},
  \bibinfo{author}{\bibfnamefont{K.~S.} \bibnamefont{Novoselov}},
  \bibnamefont{and} \bibinfo{author}{\bibfnamefont{A.~K.} \bibnamefont{Geim}},
  \bibinfo{title}{The electronic properties of graphene,}
  \bibinfo{journal}{Rev. Mod. Phys.} \textbf{\bibinfo{volume}{81}},
  \bibinfo{pages}{109} (\bibinfo{year}{2009}).

\bibitem[{\citenamefont{Murphy and R.~H.~Good}(1956)}]{MurphGood1956}
\bibinfo{author}{\bibfnamefont{E.~L.} \bibnamefont{Murphy}} \bibnamefont{and}
  \bibinfo{author}{\bibfnamefont{J.}~\bibnamefont{R.~H.~Good}},
  \bibinfo{title}{Thermionic Emission, Field Emission, and the Transition Region,}
  \bibinfo{journal}{Phys. Rev.} \textbf{\bibinfo{volume}{102}},
  \bibinfo{pages}{1464} (\bibinfo{year}{1956}).

\bibitem[{\citenamefont{X~Wei}(2013)}]{xianlongwei2013}
\bibinfo{author}{\bibfnamefont{Q.~C. L.~P.} \bibnamefont{X~Wei},
  \bibfnamefont{S~Wang}},
  \bibinfo{title}{Electron emission from a two-dimensional crystal with atomic thickness,}
  \bibinfo{journal}{AIP Advances}
  \textbf{\bibinfo{volume}{3}}, \bibinfo{pages}{042130} (\bibinfo{year}{2013}).

\bibitem[{\citenamefont{Smythe}(1950)}]{Smythe}
\bibinfo{author}{\bibfnamefont{W.~R.} \bibnamefont{Smythe}},
  \emph{\bibinfo{title}{Static and Dynamic Electricity},}
  (\bibinfo{publisher}{McGraw-Hill Book Company}, \bibinfo{address}{New York},
  \bibinfo{year}{1950}).

\bibitem[{\citenamefont{Simmons}(1963)}]{Simmons1963}
\bibinfo{author}{\bibfnamefont{J.~G.} \bibnamefont{Simmons}},
\bibinfo{title}{Generalized Formula for the Electric Tunnel Effect between Similar Electrodes Separated by a Thin Insulating Film,}
  \bibinfo{journal}{J. Appl. Phys.} \textbf{\bibinfo{volume}{34}},
  \bibinfo{pages}{1793} (\bibinfo{year}{1963}).

\bibitem[{\citenamefont{Koh and Ang}(June 2008)}]{Koh2008}
\bibinfo{author}{\bibfnamefont{W.~S.} \bibnamefont{Koh}} \bibnamefont{and}
  \bibinfo{author}{\bibfnamefont{L.~K.} \bibnamefont{Ang}},
  \bibinfo{title}{Quantum model of space–charge-limited field emission in a nanogap,}
  \bibinfo{journal}{Nanotechnology} \textbf{\bibinfo{volume}{19}},
  \bibinfo{pages}{235402} (\bibinfo{year}{June 2008}).

\bibitem[{\citenamefont{Wood}(1992)}]{Wooddc1992}
\bibinfo{author}{\bibfnamefont{D.~C.} \bibnamefont{Wood}},
  \emph{\bibinfo{title}{The Computation of Polylogarithms. Technical Report
  15-92}} (\bibinfo{publisher}{University of Kent Computing Laboratory},
  \bibinfo{address}{Canterbury, UK}, \bibinfo{year}{1992}).

\bibitem[{\citenamefont{Giovannetti et~al.}(2008)\citenamefont{Giovannetti,
  Khomyakov, Brocks, Karpan, van~den Brink, and Kelly}}]{Giovannetti2008}
\bibinfo{author}{\bibfnamefont{G.}~\bibnamefont{Giovannetti}},
  \bibinfo{author}{\bibfnamefont{P.~A.} \bibnamefont{Khomyakov}},
  \bibinfo{author}{\bibfnamefont{G.}~\bibnamefont{Brocks}},
  \bibinfo{author}{\bibfnamefont{V.~M.} \bibnamefont{Karpan}},
  \bibinfo{author}{\bibfnamefont{J.}~\bibnamefont{van~den Brink}},
  \bibnamefont{and} \bibinfo{author}{\bibfnamefont{P.~J.} \bibnamefont{Kelly}},
  \bibinfo{title}{Doping Graphene with Metal Contacts,}
  \bibinfo{journal}{Phys. Rev. Lett.} \textbf{\bibinfo{volume}{101}},
  \bibinfo{pages}{026803} (\bibinfo{year}{2008}).

\bibitem[{\citenamefont{Gan et~al.}(2013)\citenamefont{Gan, Shiue, Gao, Mak,
  Yao, Li, Szep, Walker, Hone, Heinz et~al.}}]{Englund2013}
\bibinfo{author}{\bibfnamefont{X.}~\bibnamefont{Gan}},
  \bibinfo{author}{\bibfnamefont{R.-J.} \bibnamefont{Shiue}},
  \bibinfo{author}{\bibfnamefont{Y.}~\bibnamefont{Gao}},
  \bibinfo{author}{\bibfnamefont{K.~F.} \bibnamefont{Mak}},
  \bibinfo{author}{\bibfnamefont{X.}~\bibnamefont{Yao}},
  \bibinfo{author}{\bibfnamefont{L.}~\bibnamefont{Li}},
  \bibinfo{author}{\bibfnamefont{A.}~\bibnamefont{Szep}},
  \bibinfo{author}{\bibfnamefont{D.}~\bibnamefont{Walker}},
  \bibinfo{author}{\bibfnamefont{J.}~\bibnamefont{Hone}},
  \bibinfo{author}{\bibfnamefont{T.~F.} \bibnamefont{Heinz}},
  \bibnamefont{et~al.},
  \bibinfo{title}{High-Contrast Electrooptic Modulation of a Photonic Crystal Nanocavity by Electrical Gating of Graphene,}
  \bibinfo{journal}{Nano Letters}
  \textbf{\bibinfo{volume}{13}}, \bibinfo{pages}{691} (\bibinfo{year}{2013}).

\bibitem[{\citenamefont{Petrovic et~al.}(2013)\citenamefont{Petrovic, Rakic,
  Runte, Busse, Sadowski, Lazic, Pletikosic, Pan, Milun, Pervan
  et~al.}}]{Kralj2014}
\bibinfo{author}{\bibfnamefont{M.}~\bibnamefont{Petrovic}},
  \bibinfo{author}{\bibfnamefont{I.~S.} \bibnamefont{Rakic}},
  \bibinfo{author}{\bibfnamefont{S.}~\bibnamefont{Runte}},
  \bibinfo{author}{\bibfnamefont{C.}~\bibnamefont{Busse}},
  \bibinfo{author}{\bibfnamefont{J.~T.} \bibnamefont{Sadowski}},
  \bibinfo{author}{\bibfnamefont{P.}~\bibnamefont{Lazic}},
  \bibinfo{author}{\bibfnamefont{I.}~\bibnamefont{Pletikosic}},
  \bibinfo{author}{\bibfnamefont{Z.~H.} \bibnamefont{Pan}},
  \bibinfo{author}{\bibfnamefont{M.}~\bibnamefont{Milun}},
  \bibinfo{author}{\bibfnamefont{P.}~\bibnamefont{Pervan}},
  \bibnamefont{et~al.},
  \bibinfo{title}{The mechanism of caesium intercalation of graphene,}
  \bibinfo{journal}{Nature Communications}
  \textbf{\bibinfo{volume}{4}}, \bibinfo{pages}{2772} (\bibinfo{year}{2013}).

\bibitem[{\citenamefont{Kun et~al.}(2012)\citenamefont{Xu, Zeng, Zhang, Yan,
  Ye, Wang, Seabaugh, Xing, Suehle, Richter et~al.}}]{kunxu2012nl}
\bibinfo{author}{\bibfnamefont{K.}~\bibnamefont{Xu}},
  \bibinfo{author}{\bibfnamefont{C.} \bibnamefont{Zeng}},
  \bibinfo{author}{\bibfnamefont{Q.}~\bibnamefont{Zhang}},
  \bibinfo{author}{\bibfnamefont{R.S.} \bibnamefont{Ru}},
  \bibinfo{author}{\bibfnamefont{P.}~\bibnamefont{Ye}},
  \bibinfo{author}{\bibfnamefont{K.}~\bibnamefont{Wang}},
  \bibinfo{author}{\bibfnamefont{A.}~\bibnamefont{Seabaugh}},
  \bibinfo{author}{\bibfnamefont{H.}~\bibnamefont{Xing}},
  \bibinfo{author}{\bibfnamefont{J. S.}~\bibnamefont{Suehle}},
  \bibinfo{author}{\bibfnamefont{C.~A.} \bibnamefont{Richter}},
  \bibnamefont{et~al.},
  \bibinfo{title}{Direct Measurement of Dirac Point Energy at the Graphene/Oxide Interface,}
  \bibinfo{journal}{Nano Letters}
  \textbf{\bibinfo{volume}{13}}, \bibinfo{pages}{131} (\bibinfo{year}{2012}).

\bibitem[{\citenamefont{Koch et~al.}(2014)\citenamefont{Christodoulou, Giannakopoulos, Nardi, Ligorio,
  Oehzelt, Chen, Pasquali, Timpel, Giglia, Nannarone et~al.}}]{Koch2014JPCC}
\bibinfo{author}{\bibfnamefont{C.}~\bibnamefont{Christodoulou}},
  \bibinfo{author}{\bibfnamefont{A.} \bibnamefont{Giannakopoulos}},
  \bibinfo{author}{\bibfnamefont{M.V.}~\bibnamefont{Nardi}},
  \bibinfo{author}{\bibfnamefont{G.} \bibnamefont{Ligorio}},
  \bibinfo{author}{\bibfnamefont{M.}~\bibnamefont{Oehzelt}},
  \bibinfo{author}{\bibfnamefont{L.}~\bibnamefont{Chen}},
  \bibinfo{author}{\bibfnamefont{L.}~\bibnamefont{Pasquali}},
  \bibinfo{author}{\bibfnamefont{M.}~\bibnamefont{Timpel}},
  \bibinfo{author}{\bibfnamefont{A.}~\bibnamefont{Giglia}},
  \bibinfo{author}{\bibfnamefont{S.} \bibnamefont{Nannarone}},
  \bibnamefont{et~al.},
  \bibinfo{title}{Tuning the Work Function of Graphene-on-Quartz with a High Weight Molecular Acceptor,}
  \bibinfo{journal}{J. Phys. Chem. C}
  \textbf{\bibinfo{volume}{118}}, \bibinfo{pages}{4784} (\bibinfo{year}{2014}).


\bibitem[{\citenamefont{Zhen and Song}(2011)}]{changqing2011}
\bibinfo{author}{\bibfnamefont{W.~T.}~\bibnamefont{Zheng}},
    \bibinfo{author}{\bibfnamefont{C. Q}~\bibnamefont{Song}},
    \bibinfo{title}{Underneath the fascinations of carbon nanotubes and graphene nanoribbons,}
  \bibinfo{journal}{Energy and Environmental Science}
  \textbf{\bibinfo{volume}{4}}, \bibinfo{pages}{627} (\bibinfo{year}{2011}).

\bibitem[{\citenamefont{Johannes~Voss and Abild-Pedersen}(2014)}]{voss2014}
\bibinfo{author}{\bibnamefont{Johannes~Voss},
  \bibfnamefont{Aleksandra~Vojvodic}} \bibnamefont{and}
  \bibinfo{author}{\bibfnamefont{F.}~\bibnamefont{Abild-Pedersen}},
  \bibinfo{title}{Inherent Enhancement of Electronic Emission from Hexaboride Heterostructure,}
  \bibinfo{journal}{Phys. Rev. Appl.} \textbf{\bibinfo{volume}{2}},
  \bibinfo{pages}{024004} (\bibinfo{year}{2014}).

\end{thebibliography}

\newpage

\begin{figure}
\begin{flushleft}
    \Large{CAPTIONS}
\end{flushleft}

\caption{(Color online)
(a) Schematic diagram of the electron thermionic emission process of a graphene-based thermionic energy converter (TIC), where $T_c$ and $T_a$ are the cathode temperature and anode temperature, respectively; $J_c$ and $J_a$ are current flow from cathode to anode and from anode to cathode, respectively. (b) Energy level for the TIC shown in (a). $\Phi_c$ and $\Phi_a$ are the work function of cathode and of anode, respectively. $\Phi_{eff}$ is the effective barrier height of the cathode due to Schottky lowering effect. The orange (and blue) line represents the potential profile without (with) Schottky lowering effect. The red line traces the thermionic emission process of an hot electron into vacuum.}

\caption{(Color online)
(a) The emitted current density $J$ [A/m$^2$] as a function of temperature $T$ [K] for our model (solid lines) and RD law (dashed lines) at three different work function $\Phi$ [eV] =3, 4 and 5 (top to bottom).
The inset shows $ln(J/T^3)$ (left $y$ axis) and $ln(J/T^2)$ (right $y$ axis) versus $1/T$. (b) A comparison of the analytical formula (Eq. 8) [red line] with the numerical solution of Eq. (7) [circle] as a function of temperature (for $E_f$ = 0.083 eV and $\Phi$ = 4.514 eV).}

\caption{(Color online)
(a) The current density $J$ as a function of temperature $T$ for various $E_F$ [eV]= 0.083 (intrinsic), 0.1, 0.25 and 0.4 (bottom to top). (b )$J$  as a function of $T$  at higher $E_F$ [eV] = 0.8 and 0.9, and lower $\Phi$ [eV] = 2.5 and 3. The inset presents the result at $E_F$ = 0.8 eV and $\Phi$ = 3 eV with a much lower $J$.
}

\caption{(Color online)
(a) A comparison of our calculated results (black solid line: 4.514 eV, magenta solid line: 4.66 eV) and of RD law (red) [blue dashed line: 4.66 eV, red dashed line: 4.514 eV] with experimental results \cite{jiangkaili2014} (black square) for electron emission from a single layer suspended graphene (no substrate effect). From the experiment, the emission area of graphene is $5\times 2.5$ mm$^{2}$ and the anode voltage (500 V) is separated from the graphene surface by $3.7$ mm. The applied $F$ is less 1 V/$\mu m$ so Schottky lowering effect can be ignored. (b) A comparison with experimental result \cite{Starodub2012} (symbols) and RD law (line), where the effect of substrate is important as the graphene is on top of a metallic substrate.
}

\caption{(Color online)
Efficiency of a graphene-based cathode thermionic energy converter (TIC) as a function of the work function of metallic anode $\Phi_a$ at four different anode temperature $T_{a}$ = 300, 500, 800 and 1000 K and fixed cathode temperature $T_c$ = 1800 K. The inset gives the maximum efficiency $\eta_{max}$ (left) and the required $\Phi_a$ (right) as a function of $T_a$ at $T_c$ = 900 K (black) and 1800 K (red).
}

\end{figure}
\begin{figure}[htp]
\centering
\includegraphics[scale=0.7]{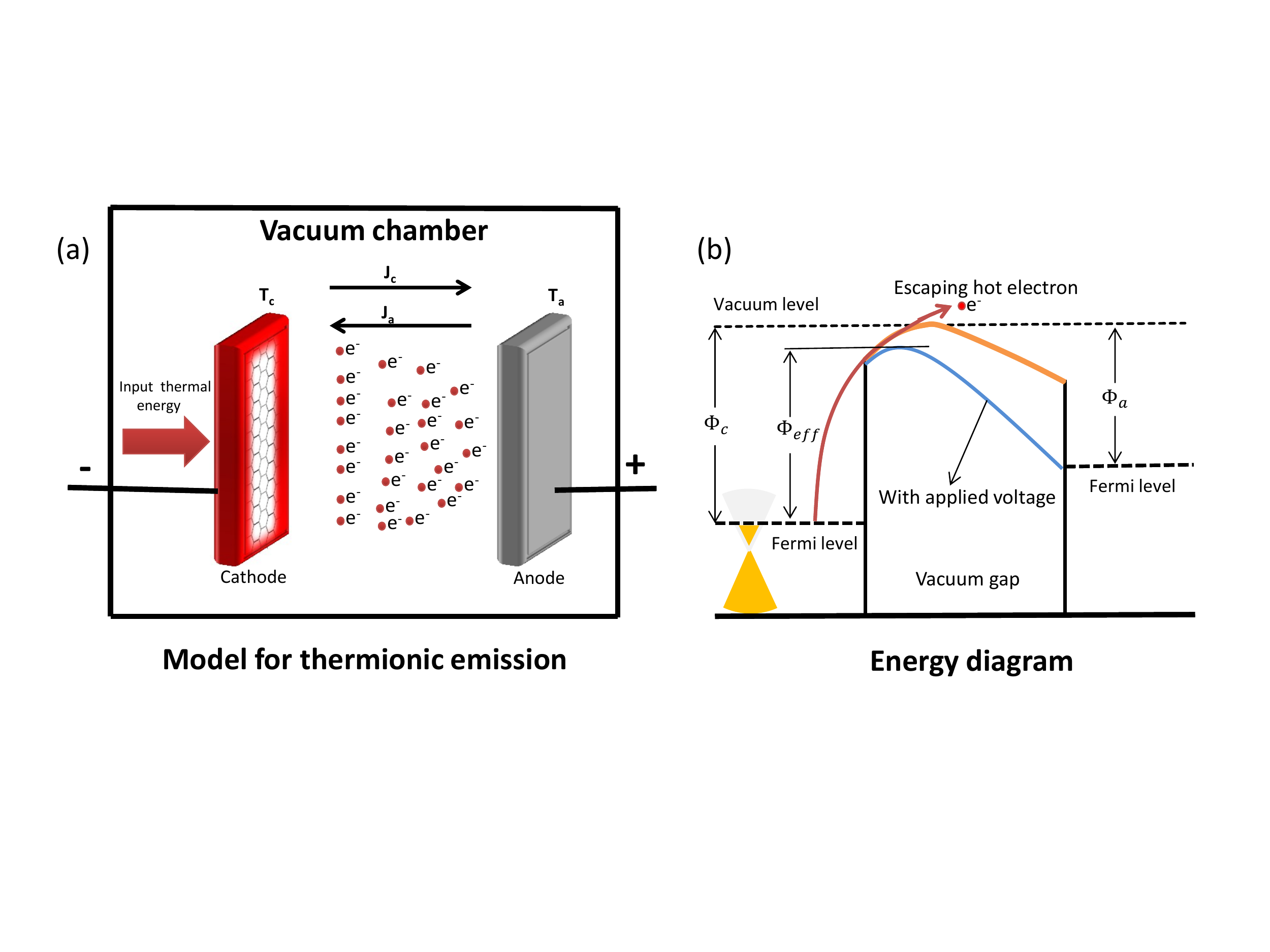}
\caption{(a) Schematic diagram of the electron thermionic emission process of a graphene-based thermionic energy converter (TIC), where $T_c$ and $T_a$ are the cathode temperature and anode temperature, respectively; $J_c$ and $J_a$ are current flow from cathode to anode and from anode to cathode, respectively. (b) Energy level for the TIC shown in (a). $\Phi_c$ and $\Phi_a$ are the work function of cathode and of anode, respectively. $\Phi_{eff}$ is the effective barrier height of the cathode due to Schottky lowering effect. The orange (and blue) line represents the potential profile without (with) Schottky lowering effect. The red line traces the thermionic emission process of an hot electron into vacuum.}
\end{figure}

\begin{figure}[htp]
\centering
\vspace{-2cm}
\includegraphics[scale=0.5]{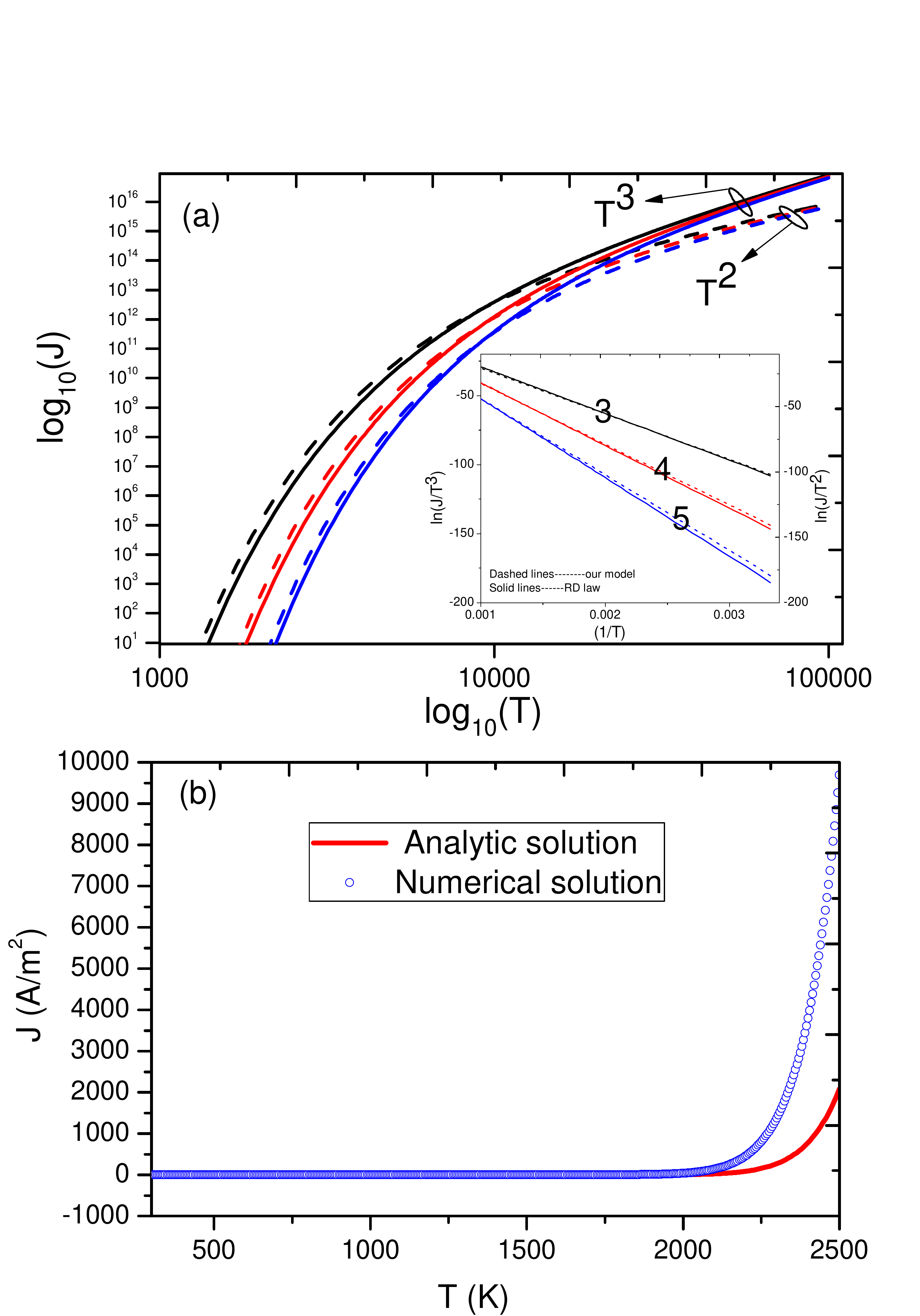}
\caption{(a) The emitted current density $J$ [A/m$^2$] as a function of temperature $T$ [K] for our model (solid lines) and RD law (dashed lines) at three different work function $\Phi$ [eV] =3, 4 and 5 (top to bottom).
The inset shows $ln(J/T^3)$ (left $y$ axis) and $ln(J/T^2)$ (right $y$ axis) versus $1/T$. (b) A comparison of the analytical formula (Eq. 8) [red line] with the numerical solution of Eq. (7) [circle] as a function of temperature (for $E_f$ = 0.083 eV and $\Phi$ = 4.514 eV).}
\end{figure}

\begin{figure}[htp]
\centering
\includegraphics[scale=0.5]{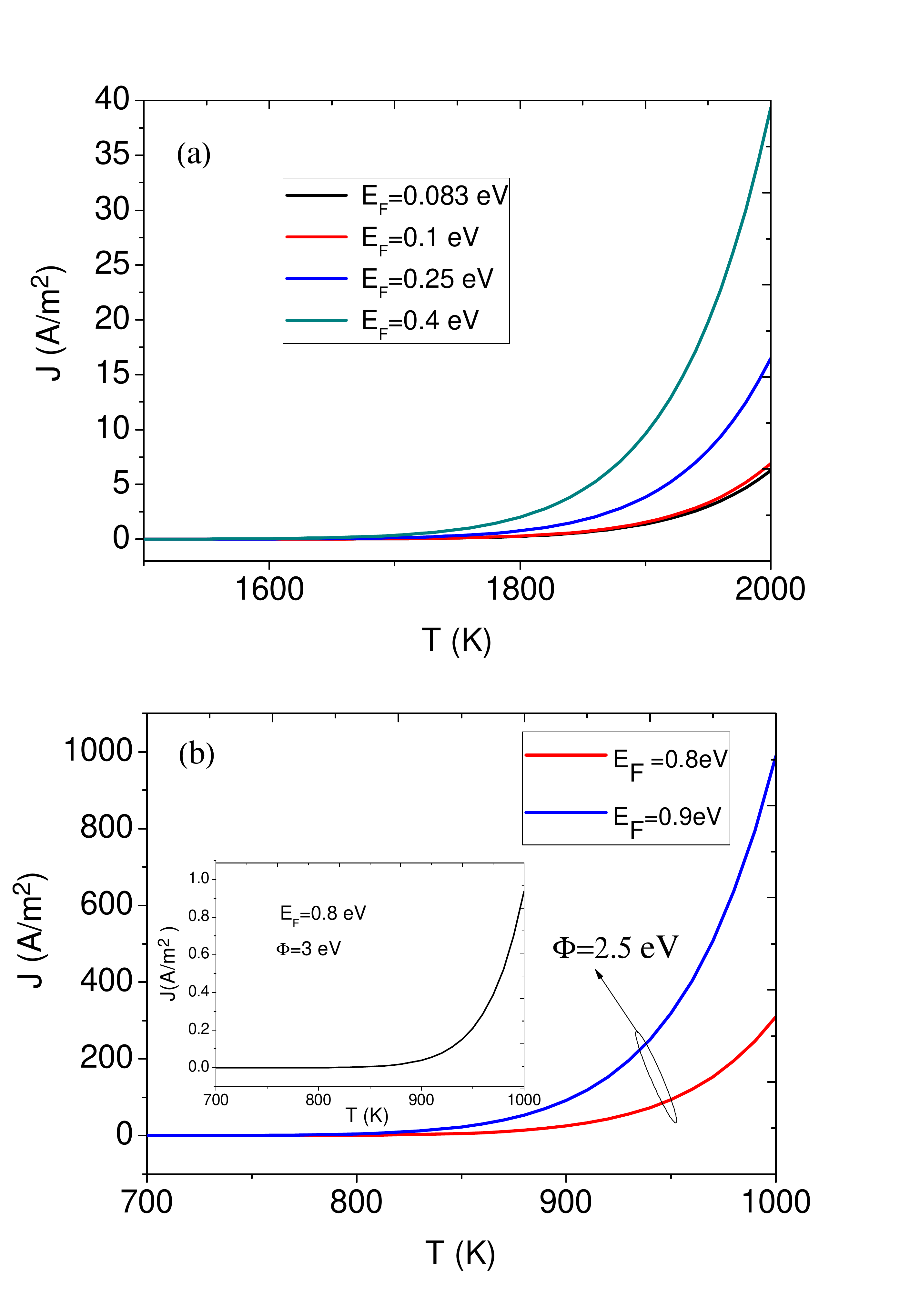}
\caption{The current density $J$ as a function of temperature $T$ for various $E_F$ [eV]= 0.083 (intrinsic), 0.1, 0.25 and 0.4 (bottom to top). (b )$J$  as a function of $T$  at higher $E_F$ [eV] = 0.8 and 0.9, and lower $\Phi$ [eV] = 2.5 and 3. The inset presents the result at $E_F$ = 0.8 eV and $\Phi$ = 3 eV with a much lower $J$.}
\end{figure}

\begin{figure}[htp]
%\centering
\hspace{-4cm}
\includegraphics[scale=0.8]{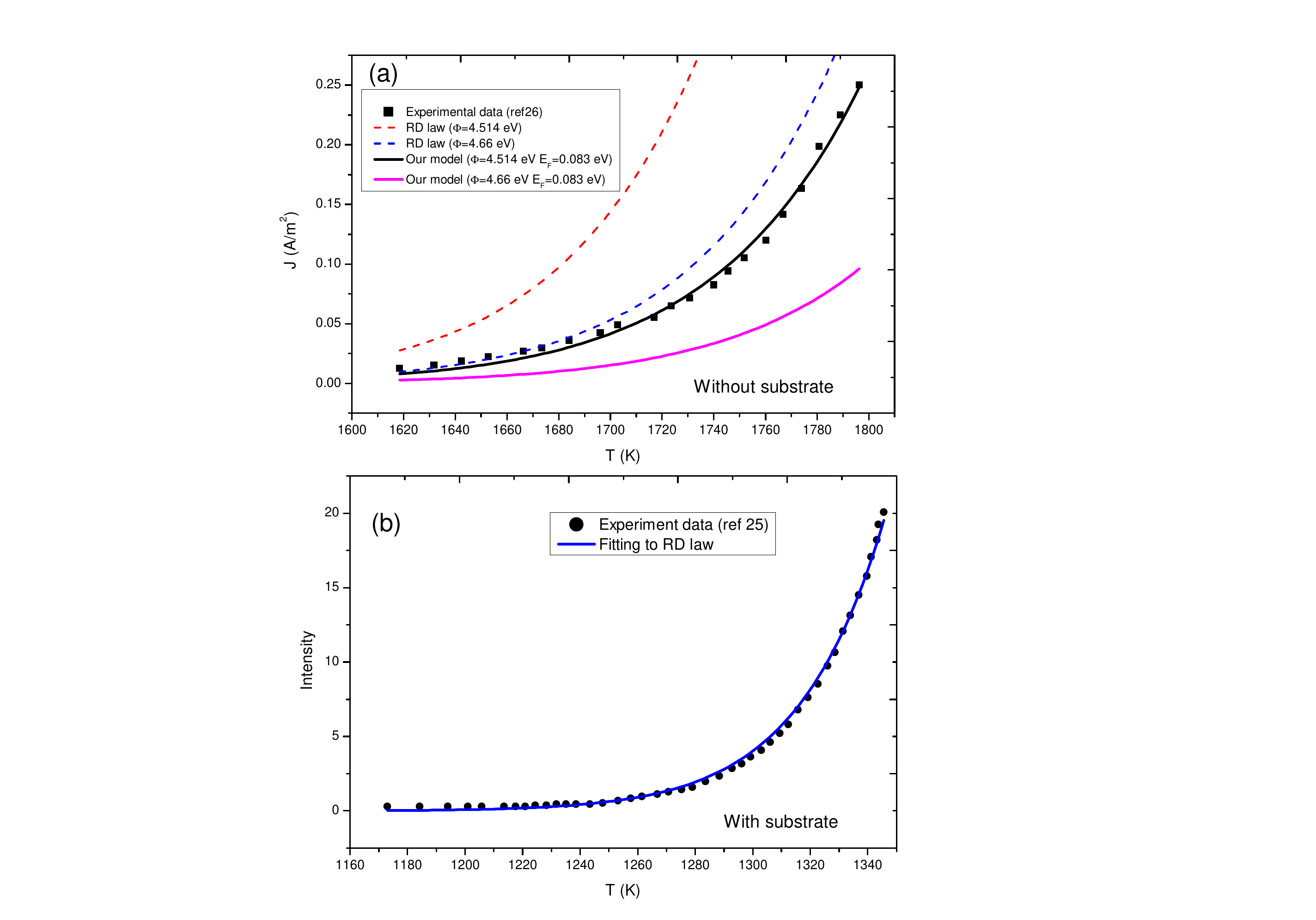}
\caption{(a) A comparison of our calculated results (black solid line: 4.514 eV, magenta solid line: 4.66 eV) and of RD law (red) [blue dashed line: 4.66 eV, red dashed line: 4.514 eV] with experimental results \cite{jiangkaili2014} (black square) for electron emission from a single layer suspended graphene (no substrate effect). From the experiment, the emission area of graphene is $5\times 2.5$ mm$^{2}$ and the anode voltage (500 V) is separated from the graphene surface by $3.7$ mm. The applied $F$ is less 1 V/$\mu m$ so Schottky lowering effect can be ignored. (b) A comparison with experimental result \cite{Starodub2012} (symbols) and RD law (line), where the effect of substrate is important as the graphene is on top of a metallic substrate.}
\end{figure}

\begin{figure}[htp]
\centering
\includegraphics[scale=0.7]{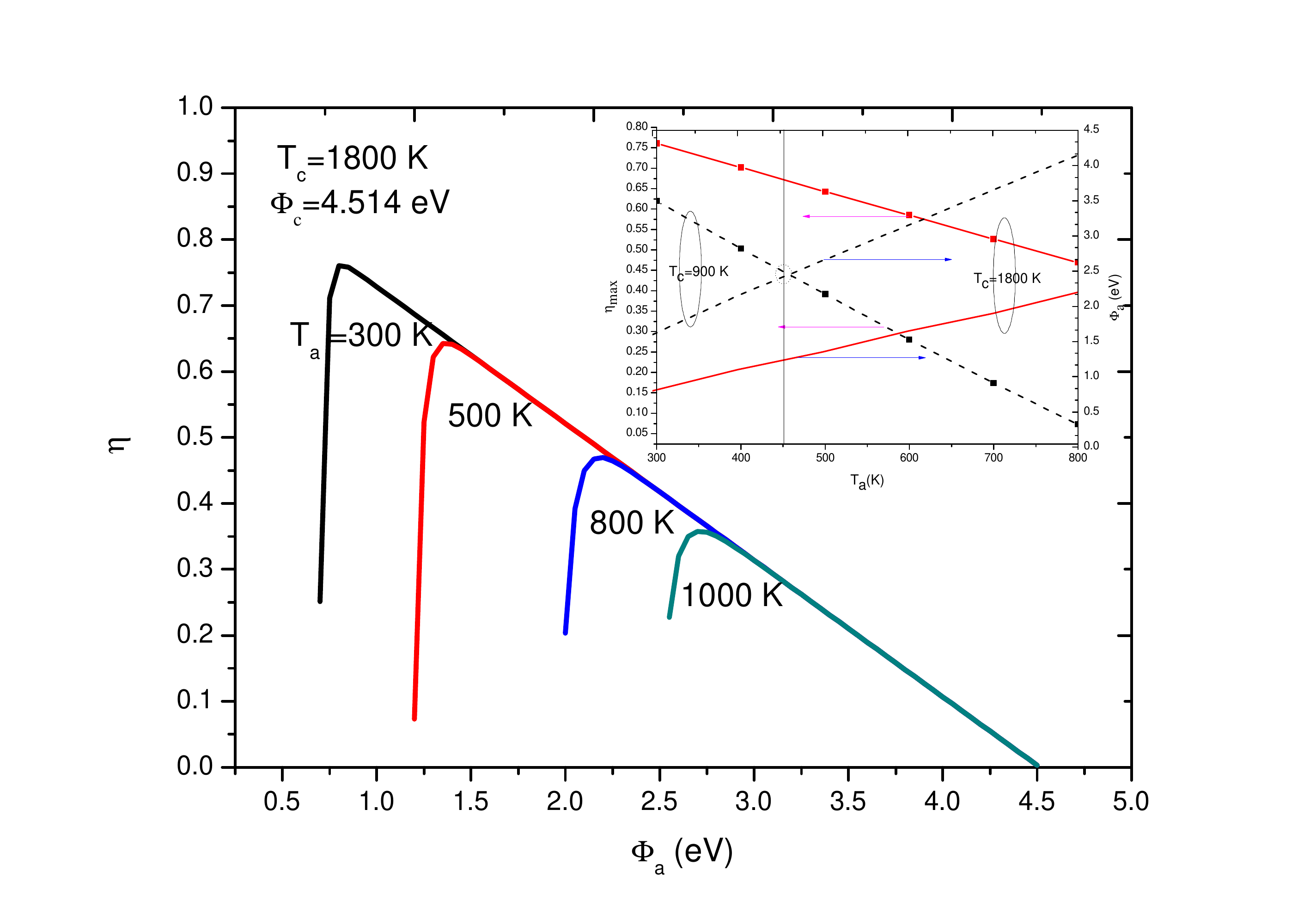}
\caption{Efficiency of a graphene-based cathode thermionic energy converter (TIC) as a function of the work function of metallic anode $\Phi_a$ at four different anode temperature $T_{a}$ = 300, 500, 800 and 1000 K and fixed cathode temperature $T_c$ = 1800 K. The inset gives the maximum efficiency $\eta_{max}$ (left) and the required $\Phi_a$ (right) as a function of $T_a$ at $T_c$ = 900 K (black) and 1800 K (red).}
\end{figure}

\end{document}